\documentstyle[12pt,epsfig]{article}

\begin{document}
\def\Journal#1#2#3#4{#1 {\bf #3} (#2), #4}
\def\NCA{\em Nuovo Cimento}
\def\NIM{\em Nucl. Instrum. Methods}
\def\NIMA{{\em Nucl. Instrum. Methods} A}
\def\npbs{{\em Nucl. Phys.} B}
\def\plb{{\em Phys. Lett.}  B}
\def\prl{{\em Phys. Rev. Lett.}}
\def\prd{{\em Phys. Rev.} D}
\def\zpc{{\em Z. Phys.} C}
\def\epj{{\em Eur. Phys. J.} C}
\def\p{\mbox{$p$}}
\def\bar{\overline}
\newcommand {\bb} {\ifmmode {b\bar{b}}\else {$b\bar{b}$}\fi}
\newcommand {\cc} {\ifmmode {c\bar{c}}\else {$c\bar{c}$}\fi}
\newcommand {\uds}{\ifmmode {uds}\else {$uds$}\fi}
\newcommand {\ra} {\ifmmode {\rightarrow}\else {$\rightarrow$}\fi}
\newcommand {\bc} {\begin{center}}
\newcommand {\bt} {\begin{table}}
\newcommand {\et} {\end{table}}
\newcommand {\ec} {\end{center}}
\newcommand {\Rb} {\ifmmode {\mathrm{R}_b} \else $\mathrm{R}_b$ \fi }
\newcommand {\fd} {\ifmmode {\mathrm{f}_d} \else $\mathrm{f}_d$ \fi }
\newcommand {\als} {\mbox{$\alpha_S$}}
\newcommand {\alphas} {\mbox{$\alpha_S(M_{\rm{Z}^0})$}}
\newcommand {\alphasmu} {\mbox{$\alpha_S(\mu)$}}
\newcommand {\eec}{\mbox{$\Sigma_{EEC}$}}
\newcommand {\eecchi}{\mbox{$\Sigma_{EEC}(\chi)$}}
\newcommand {\aeec}{\mbox{$\Sigma_{AEEC}$}}
\newcommand {\aeecchi}{\mbox{$\Sigma_{AEEC}(\chi)$}}
\newcommand {\scalef}{\mbox{$x_{\mu}$}}
\newcommand {\Ds} {\ifmmode {\mathrm{D}^{\star +}} \else {D$^{\star +}$} \fi}
\newcommand {\Dz} {\ifmmode {\mathrm{D}^{0}} \else {D$^{0}$} \fi}
\newcommand {\bl}    {BR({\rm{b \rightarrow \ell}})}
\newcommand {\cl}    {BR({\rm{c \rightarrow \ell}})}
\newcommand {\bcbl}   {BR({\rm{b \rightarrow \bar{c} \rightarrow \ell}})}
\newcommand {\bcl}   {BR({\rm{b \rightarrow c \rightarrow \ell}})}
\newcommand {\btaul}  {BR({\rm{b \rightarrow \tau \rightarrow \ell}})}
\newcommand {\bpsill} {BR({\rm{b \rightarrow J/\psi\rightarrow \ell^+\ell^-}})}
\newcommand {\glcc}   {\rm{g \rightarrow c \bar c}}
\newcommand {\glbb}   {\rm{g \rightarrow b \bar b}}
\newcommand {\bu} {\rm {b\ra u \ra \ell^+}}
\def\vub{$|V_{ub}|$}
\def\GeV{{\mbox{\mathrm GeV}}}
\def\Dstar{\ifmmode {{\mathrm D}^*} \else {${\mathrm D}^*$} \fi}
\def\Dzero{\ifmmode {{\mathrm D}^0} \else {${\mathrm D}^0$} \fi}
\def\md0c{M_{{\mathrm D}^0}^{\mathrm cand}}
\def\etal{{\it et al.}}
\def\bsbar{${\rm \overline{B_s^0}}$}
\def\bsbs{${\rm B_s^0}$}
\def\bd{${\rm B^0_d}$}
\def\bu{${\rm B^-_u}$}
\def\bdbar{${\rm {\overline B}_d^0}$}
\def\bbar{${\rm {\overline B}^0}$}
\def\dmd{$\Delta m_{\rm d}$}
\def\dstara{${\rm D^{\star \pm}}$}
\def\dstarp{${\rm D^{\star +}}$}
\def\dsp{${\rm D^{\star +}}$}
\def\dstarm{${\rm D^{\star -}}$}
\def\dstar{${\rm D^{\star}}$}
\def\epem{${\rm e^+e^-}$}
\def\vcb{\mbox{$|V_{\rm cb}|$}}
\def\fw{${\cal F}(w)$}
\def\fone{${\cal F}(1)$}
\def\fvcb{${\cal F}(1)|V_{\rm cb}|$}
\def\btods{${\rm \overline B_d}^0\to {\rm D}^{*+}\ell^-{\overline \nu}_\ell$}
\def\btodss{$\rm b \to D^{\star \star}\ell^-{\overline \nu}_\ell$}
\newcommand {\Bz} {\ifmmode {\rm \bar{B}^0_d} \else {$\rm \bar{B}^0_d$} \fi}
\newcommand {\Dss} {\ifmmode {\mathrm{D}^{**}} \else {D$^{**}$} \fi}
\newcommand {\BtoD} {\ifmmode {\Bz \to \Ds \ell \bar{\nu}}
                \else ${\Bz \to \Ds \ell \bar{\nu}}$ \fi }
\newcommand {\Btau} {\ifmmode {\bar{\rm{B}} \to \tau \bar{\nu_\tau} \Ds}
                \else ${\bar{\mathrm{B}} \to \tau \bar{\nu_\tau} \Ds }$ \fi }
\newcommand {\Bxc} {\ifmmode {\bar{\mathrm{B}} \to \mathrm{X}_c \Ds }
                \else ${\bar{\mathrm{B}} \to \mathrm{X}_c \Ds }$ \fi }
\def\dec{\rightarrow}

\begin{titlepage}

\title{ Present status of experimental determination of $|V_{cb}|$\footnote{An abbreviated version
will be included in PDG2002 as a minireview on $|V_{cb}|$. }}

\author{Marina Artuso$^1$ and
Elisabetta Barberio$^2$} \maketitle

\centerline{$^1$Syracuse University, Syracuse, NY 13244, USA}
\centerline{$^2$CERN, Geneva, Switzerland}

\begin{abstract}

The present status of our knowledge of the magnitude of the quark
mixing parameter $\rm |V_{cb}|$ is reviewed, with particular
emphasis on the factors affecting experimental and theoretical
errors and on prospects for a more precise determination.

\end{abstract}
\end{titlepage}

\section{Introduction}

In the framework of the Standard Model, the quark sector is
characterized by a rich pattern of flavor-changing transitions,
described by the Cabibbo-Kobayashi-Maskawa (CKM) matrix:
\begin{equation}
V_{CKM} =\left(\begin{array}{ccc}
V_{ud} &  V_{us} & V_{ub} \\
V_{cd} &  V_{cs} & V_{cb} \\
V_{td} &  V_{ts} & V_{tb} \end{array}\right).
\end{equation}
Since the CKM matrix must be unitary, it can be expressed as a
function of only four parameters. A commonly used approximate
parameterization was originally proposed by Wolfenstein
\cite{wolf-prl}. It reflects the hierarchy between the magnitude
of matrix elements belonging to different generations. Very
frequently it is quoted in the approximation valid only to
$\lambda^3$. We need to carry out this expansion further in order
to incorporate  CP violation in neutral $K$ decays. This
expression, accurate to $\lambda^3$ for the real part and
$\lambda^5$ for the imaginary part, is given by:
\begin{equation}
\left({\begin{array}{ccc}
1-\lambda^2/2 &  \lambda & A\lambda^3(\rho-i\eta (1-\lambda^2/2)) \\
-\lambda &  1-\lambda^2/2-i\eta A^2\lambda^4 & A\lambda^2(1+i\eta\lambda^2) \\
A\lambda^3(1-\rho-i\eta) &  -A\lambda^2& 1
\end{array}}\right).
\end{equation}
The parameter $\lambda$ is well measured as 0.2196$\pm$0.0023
\cite{PDG2002}, constraints exist on $\rho$ and $\eta$ from
measurements of $V_{ub}$ and $B^0\bar{B}^0$ mixing. This report
focuses on the magnitude of the CKM element  $\rm |V_{cb}|$,
related to the Wolfenstein parameter $A$ \cite{wolf-prl}.

 Two different methods have
been used to extract this parameter from data: the {\bf exclusive}
measurement, where \vcb\ is extracted by studying exclusive $\rm
B\dec D^{\star}\ell \nu$ and $\rm B\dec D \ell \nu$ decay
processes; and the {\bf inclusive} measurement, which uses the
semileptonic width of b-hadron decays. Theoretical estimates play
a crucial role in extracting $\rm |V_{cb}|$ and an understanding
of their uncertainties is very important.

\section{Exclusive \vcb\ determination}

The exclusive \vcb\ determination is obtained studying the
$\rm B\dec D^{\star} \ell \nu$ and $\rm B\dec D \ell \nu$ decays,
using Heavy Quark Effective Theory (HQET), an exact theory in the limit
of infinite quark masses.
Presently the $\rm B\dec D \ell \nu$ transition
provides a less precise value and is used as a check.

\subsection{The decay $\rm B\dec D^{\star} \ell \nu$ in HQET}

HQET predicts that the differential partial decay width for
this process, $d\Gamma/dw$, is related to \vcb\ through:
\begin{equation}
\frac{d\Gamma}{dw}(B\rightarrow D^{\star}\ell \nu) = \frac{G_F^2
|V_{cb}|^2}{48\pi^3}{\cal K}(w){\cal F}(w)^2,
\end{equation}
where $w$ is the inner product of the $B$ and $D^{\star}$ meson 4-velocities,
${\cal K}(w)$ is a known phase space factor and the form factor ${\cal F}(w)$
is generally expressed as the product of a normalization factor ${\cal F}(1)$
and a function, $g(w)$, constrained by dispersion relations
\cite{grinstein}.

There are several different corrections to the infinite mass value
${\cal F}(1)=1$ \cite{babarph}:
\begin{equation}
{\cal F}(1) =\eta _{QED}\eta _A \left[ 1 + \delta _{1/m_Q^2} +
...\right]
\end{equation}
where $Q=c$ or $b$. By virtue of Luke's theorem \cite{luke}, the
first term in the non-perturbative expansion in powers of $1/m_Q$
vanishes. $QED$ corrections up to leading logarithmic order give
$\eta_{QED}\approx 1.007$ \cite{babarph} and $QCD$ radiative
corrections to two loops give $\eta _A = 0.960\pm 0.007$
\cite{chan}. Different estimates of the $1/m_Q^2$ corrections,
involving terms proportional to $1/m_b^2$, $1/m_c^2$ and
$1/(m_bm_c)$ have been performed in a quark model
\cite{neubert-vcb}, \cite{neubert-vcb2}, with OPE sum rules
\cite{uraltsev-vcb}, and, more recently, with an HQET based
lattice gauge calculation \cite{kronfeld}. The value from this
quenched lattice HQET calculation is $0.913^{+0.024}_{-0.017}\pm
0.016^{+0.003}_{-0.014}~^{+0.000}_{-0.016} ~^{+0.006}_{-0.014}$.
The errors quoted reflect the statistical accuracy, the matching
error, the lattice finite size, the uncertainty in the quark
masses and an estimate of the error induced by the quenched
approximation, respectively. The central value obtained with OPE
sum rules is similar, with an error of $\pm 0.04$
\cite{ckm-workshop}. Consequently, we will use ${\cal F}(1)=\ 0.91
\pm 0.04$ \cite{ckm-workshop}.

The analytical expression of ${\cal F}(w)$ is not known a-priori,
and this introduces an additional uncertainty in the determination
of ${\cal F}(1)|V_{cb}|$. First measurements of $\rm |V_{cb}|$
were performed assuming a linear approximation for ${\cal F}(w)$.
It has been shown \cite{Stone} that this assumption is not
justified and that linear fits systematically underestimate the
extrapolation at zero recoil ($w=1$) by about 3\%. Most of this
effect is related to the curvature of the form factor, and does
not depend strongly upon the details of the non-linear shape
chosen \cite{Stone}. All recent published results use a non-linear
shape for ${\cal F}(w)$, approximated with an expansion around
$w=1$ \cite{CLN}. ${\cal F} (w)$ is parameterized in terms of the
variable $\rho^2$, which is the slope of the form-factor at zero
recoil of Ref. \cite{CLN} .

\subsection{Experimental techniques to study the decay
$\rm B\dec D^{\star}\ell \nu$.}

The decay $\rm B\dec D^{\star}\ell \nu$ has been studied in
experiments performed at the $\Upsilon \rm(4S)$ center of mass
energy and at the $\rm Z^0$ center of mass energy at LEP. At the
$\Upsilon \rm(4S)$, experiments have the advantage that the $w$
resolution is quite good. However, they have more limited
statistics near $w=1$ in the decay $\rm \bar{B}^o\rightarrow
D^{\star +} \ell \nu$, because of the lower reconstruction
efficiency of the slow pion, from the $\rm D^{\star +} \to \pi^{+}
D^0$ decay. The decay $\rm B^-\rightarrow D^{\star 0} \ell
\bar{\nu}$ is not affected by this problem and CLEO
\cite{cleo-vcb} uses both channels. In addition, kinematic
constraints enable $\rm \Upsilon(4S)$ experiments to identify the
final state including the $\rm D^{\star}$ without a large
contamination from the poorly known semileptonic decays including
a hadronic system heavier than  $\rm D^{\star}$, commonly
identified as `$ \rm D^{\star\star}$'. At LEP, B's are produced
with a large momentum (about ~30 GeV on average). This makes the
determination of $w$ dependent upon the neutrino four-momentum
reconstruction, thus giving a relatively poor resolution and
limited physics background rejection capabilities. By contrast,
LEP experiments benefit from an efficiency only mildly dependent
upon $w$.

\begin{figure}
\centerline{\epsfig{figure=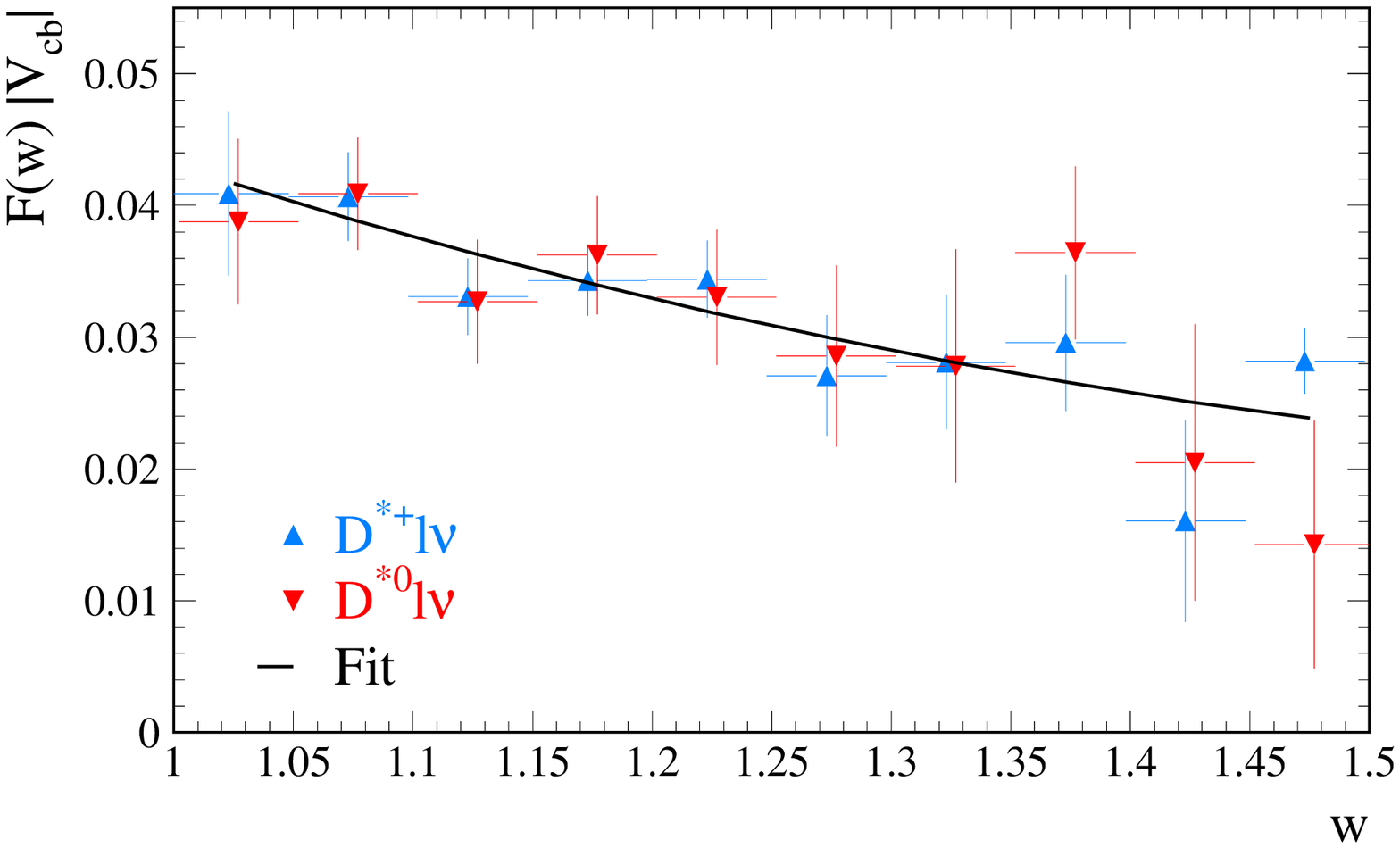,width=3.8in}}
\caption{$|V_{cb}|{\cal F}(w)$ CLEO unfolded spectra, where the
solid circles (squares) are derived from the $D^{\star +}\ell\nu$
($D^{\star o} \ell \nu$) data samples respectively. The curve
shows the result of the fit described in the text.}
\label{dslnu-cleo}
\end{figure}

\begin{figure}[htb]
\centerline{\epsfig{figure=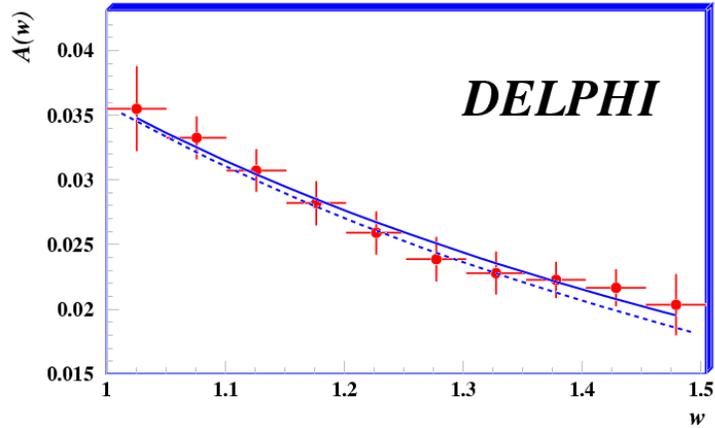,width=3.8in}}
\caption{\label{dd} a) Decay form factor from the unfolded DELPHI
data. The dotted line shows the fit result neglecting the
bin-to-bin correlation. The continuous line shows the result when
the correlation are taken into account.}
\end{figure}

Experiments determine the product $({\cal F}(1)\cdot |V_{cb}|)^2$
by fitting the measured ${d\Gamma}/{dw}$ distribution.
Measurements at the $\Upsilon(4S)$ have been performed by CLEO
\cite{cleo-vcb}  and Belle \cite{belle-dslnu}.
Figure~\ref{dslnu-cleo} shows the latest CLEO measurement
\cite{cleo-vcb} of ${\cal F}(w)| V_{cb} |$ as a function of $w$.
At LEP data are available from ALEPH \cite{ALEPH_vcb}, DELPHI
\cite{DELPHI_vcb} and OPAL \cite{OPAL_vcb}. DELPHI fits for ${\cal
F}(w)|V_{cb}|$, including a free-curvature parameter; the result
of this fit, Fig.~\ref{dd}, agrees with the shape parameterization
of \cite{CLN}, used by all experiments. Both the CLEO and DELPHI
spectra are corrected for smearing, as well as for efficiency.

At LEP, the dominant source of systematic error is the uncertainty
on the contribution to ${d\Gamma}/{dw}$ from semileptonic B decays
with final states including a hadron system heavier than the
$D^{\star}$, either narrow orbitally excited charmed meson or
non-resonant or broad species. The existence of narrow resonant
states is well established \cite{PDG2002} and a signal of a broad
resonance has been seen by CLEO \cite{CB}, but the decay
characteristics of these states in b-hadron semileptonic decays
have large uncertainties. The average of ALEPH
\cite{aleph-d2star}, CLEO \cite{cleo-d2star} and DELPHI
\cite{delphi-d2star} narrow state branching fractions show that
the ratio $
  \rm R_{\star\star} = \frac{{\cal B}(\bar{B} \to D^\star_2 \ell
\bar{\nu})} {{\cal B}(\bar{B} \to D_1 \ell \bar{\nu})}
$ is smaller than one ($<0.6$ at 95\% C.L.\cite{HFnote}), in
disagreement with HQET models where an infinite quark mass is
assumed \cite{wrong}, but in agreement with models which take into
account finite quark mass corrections \cite{ligeti}. Hence, LEP
experiments use the treatment of narrow $\rm D^{\star\star}$
proposed in \cite{ligeti}, which accounts for ${\cal O}(1/m_c)$
corrections. Ref. \cite{ligeti} provides several possible
approximations of the form factors, that depend on five different
expansion schemes and on three input parameters. To calculate the
systematic errors each proposed scheme is tested, with the
relevant input parameters varied over a range consistent with the
experimental limit on $\rm R_{\star\star}$. The quoted systematic
error is the maximal difference from the central value obtained
with this method. Broad resonances or other non-resonant terms may
not be modelled correctly with this approach.
\begin{table}
\begin{center}
\begin{tabular}{|l|c|c|c|} \hline
experiment & \fvcb\ $(\times 10^{3})$ & $\rho^2$ & $\rm
Corr_{stat}$ \\ \hline
 ALEPH published &  31.9$\pm$  1.8$\pm$ 1.9 & 0.31$\pm$ 0.17$\pm$ 0.08 & 92\% \\
 ALEPH update    &  31.5$\pm$  2.1$\pm$ 1.3 & 0.58$\pm$ 0.25$\pm$ 0.11 & 94\% \\
 DELPHI          &  35.5$\pm$  1.4$\pm$ 2.4 & 1.34$\pm$ 0.14$\pm$ 0.23 & 94\% \\
 OPAL            &  37.1$\pm$  1.0$\pm$ 2.0 & 1.21$\pm$ 0.12$\pm$ 0.20 & 90\% \\
 Belle           &  35.8$\pm$  1.9$\pm$ 1.8 & 1.45$\pm$ 0.16$\pm$ 0.20 & 90\% \\
 CLEO            &  43.1$\pm$  1.3$\pm$ 1.8 & 1.61$\pm$ 0.09$\pm$ 0.21 & 86\% \\
\hline
\end{tabular}
\end{center}
\caption{Experimental results as published by the collaborations.
LEP numbers use theoretical predictions for R$_1$ and R$_2$. The
published ALEPH result is obtained using a linear fit and the old
Isgur-Wise model \cite{isgw} for $\rm D^{\star\star}$. The updated
ALEPH numbers (used in our average) are obtained using the same
fit parameterization and $\rm D^{\star\star}$ models as the other
LEP experiments \cite{lepvcb}. The Belle result quoted here uses
R$_1$ and R$_2$ from CLEO data. \label{t:publ}}
\end{table}
\begin{table}
\begin{center}
\begin{tabular}{|l|c|c|}
\hline
 Parameter & Value & Reference \\ \hline
 $\rm \Rb=\Gamma(Z\to b\bar{b})/\Gamma(Z\to had)$ & (21.64 $\pm$ 0.07)\% & \cite{EWWG}  \\
 $\rm \fd = {\cal B}(b\to B_d)$ & (40.0  $\pm 1.1$)\% & \cite{Bosc} \\
$\tau(B^0)$& (1.54   $\pm$ 0.015) ps & \cite{life} \\
$\rm {x_E}^{LEP}=E(B~meson)/\sqrt{s}$  & 0.702  $\pm$ 0.008 & \cite{EWWG} \\
$ {\cal B}(D^{\star +}\to\Dz\pi^+$) & (67.7 $\pm$ 0.5) \% & \cite{PDG2002} \\
\hline
R$_1$ & $1.18 \pm 0.32$ & \cite{r1r2cleo} \\
R$_2$ & $0.71 \pm 0.23$ & \cite{r1r2cleo} \\ \hline
 ${\cal B}$(\Btau)           & (1.27  $\pm$ 0.21)\% &  \cite{HFnote} \\
 ${\cal B}(B^- \to D^{\star +} \pi^- \ell \bar{\nu}$) & (1.29$\pm0.16$) \% &
             \cite{HFnote}  \\
${\cal B}(\Bz~ \to~ D^{\star +} \pi^0 \ell \bar{\nu})$ &
$(0.61\pm0.08)$\% &
 \cite{HFnote} \\
${\cal B}(\rm B_s \to \Ds K \ell \bar{\nu})$ & $(0.65\pm0.23)$\% &
\cite{HFnote}
\\ \hline
\end{tabular}
\end{center}
\caption{Values of the most relevant parameters affecting the
measurement of $\vcb$. The three $\rm D^{\star\star}$ production
rates are fully correlated.} \label{tab:summary}
\end{table}

\begin{figure}
\centerline{\epsfig{figure=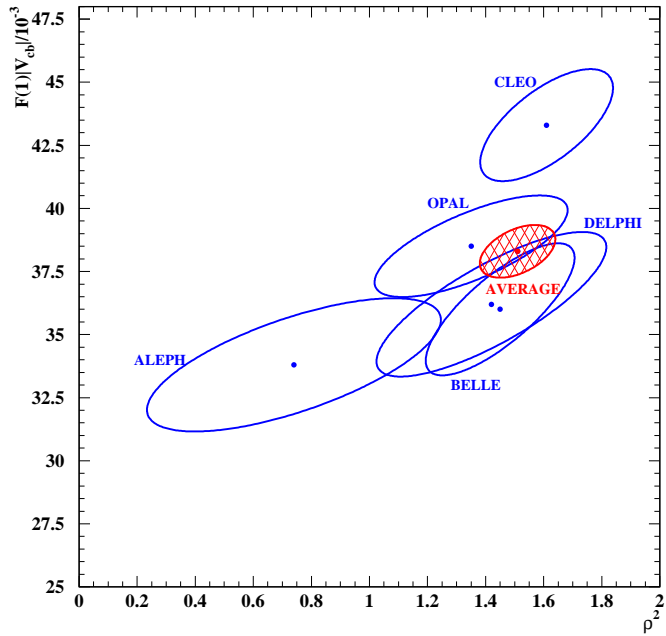,width=3.8in}}
\caption{\label{f:vcbell} The error ellipses for the corrected
measurements and world average for $|{\cal F}(1)V_{cb}|$ versus
$\rho^2$. The ellipses are obtained from product between the 1
$\sigma$ error of $|{\cal F}(1)V_{cb}|$, $\rho^2$ and the
correlation between the two. Consequently, the ellipses correspond
to about 37\% C.L.}
\end{figure}

Table~\ref{t:publ} summarizes all published data as quoted in the
original papers. 
To combine the published data, the central values and the errors
of ${\cal F}(1) \vcb$ and $\rho^2$ are re-scaled to the same set
of input parameters and their quoted uncertainties. These common
inputs are listed in Table~\ref{tab:summary}. The ${\cal
F}(1)|V_{cb}|$ values used for this average are extracted using
the parametrization in \cite{cleo-vcb}, based on the experimental
determinations of the vector and axial form factor ratios R$_1$
and R$_2$ \cite{r1r2cleo}. The LEP data, which originally used
theoretical values for these ratios, are re-scaled accordingly
\cite{lepvcb}. Table~\ref{t:correxp} summarized the corrected
data. The averaging procedure \cite{lepvcb} takes into account
statistical and systematic correlations between ${\cal F}(1) \vcb$
and $\rho^2$. Averaging the measurements in Table \ref{t:correxp},
we get:
$${\cal F} (1) \vcb = (38.3  \pm 1.0) \times 10^{-3}$$ and
$$\rho^2 = 1.5 \pm 0.13$$ with a confidence level \footnote{
The $\chi^2$ per degree of freedom is less than 2, and we do not
scale the error.} of 5.1\%. The error ellipses for the corrected
measurements and for the world average are shown in
Figure~\ref{f:vcbell}.
\begin{table}
\begin{center}
\begin{tabular}{|l|ccc|} \hline
experiment   & \fvcb\ $(\times 10^{3})$ & $\rho^2$ & $\rm
Corr_{stat}$ \\ \hline
 ALEPH       &  33.8$\pm$  2.1$\pm$ 1.6 & 0.74$\pm$ 0.25$\pm$ 0.41  & 94\% \\
 DELPHI      &  36.1$\pm$  1.4$\pm$ 2.5 & 1.42$\pm$ 0.14$\pm$ 0.37 & 94\% \\
 OPAL        &  38.5$\pm$  0.9$\pm$ 1.8 & 1.35$\pm$ 0.12$\pm$ 0.31 &  89\% \\
 Belle       &  36.0$\pm$  1.9$\pm$ 1.8 & 1.45$\pm$ 0.16$\pm$ 0.20 &  90\% \\
 CLEO        &  43.3$\pm$  1.3$\pm$ 1.8 & 1.61$\pm$ 0.09$\pm$ 0.21 &  86\% \\
\hline
World average&  38.3 $\pm$ 0.5$\pm$ 0.9 & 1.51$\pm$ 0.05$\pm$ 0.12 & 86\%\\
\hline
\end{tabular}
\end{center}
\caption{Experimental results after the correction to common
inputs and world average. The LEP numbers are corrected to use
R$_1$ and R$_2$ from CLEO data. $\rho^2$ is the slope of the
form-factor at zero recoil as defined in \cite{CLN}.
\label{t:correxp}}
\end{table}

The main contributions to the ${\cal F} (1) \vcb$ systematic error
are from the uncertainty on the $\rm B\dec D^{\star\star}\ell \nu$
shape and $\rm {\cal B}(\rm b\to B_d)$, ($0.57\times 10^{-3}$),
fully correlated among the LEP experiments, the branching fraction
of $\rm D$ and $\rm D^{\star}$ decays,($0.4\times 10^{-3}$), fully
correlated among all the experiments, and the slow pion
reconstruction from Belle and CLEO which are uncorrelated,
(0.28$\times 10^{-3}$). The main contribution to the $\rho^2$
systematic error is from the uncertainties in the measured values
of R$_1$ and R$_2$ (0.12), fully correlated among experiments.
Because of the large contribution of this uncertainty to the
non-diagonal terms of the covariance matrix, the averaged $\rho^2$
is higher than one would naively expect.

Using ${\cal F} (1) = 0.91 \pm 0.04$ \cite{ckm-workshop}, we get
$\vcb = ( 42.1 \pm 1.1_{exp} \pm 1.9_{theo}) \times 10^{-3}$. The
dominant error is theoretical, but there are good prospects that
lattice gauge calculations will improve significantly the accuracy
of their estimate.
\subsection{The decay $\rm B\dec D \ell \nu$}

The study of the decay $\rm B\dec D \ell \nu$ poses new challenges
both from the theoretical and experimental point of view.

The differential decay rate for $\rm B\dec D \ell \nu$ can be
expressed as:
\begin{equation}
\frac{d\Gamma_D}{dw} (B\dec D\ell\nu)= \frac{G_F^2|
V_{cb}|^2}{48\pi^3}{\cal K_D}(w){\cal G}(w)^2,
\end{equation}
where $w$ is the inner product of the B and D meson 4-velocities,
${\cal K_D}(w)$ is the phase space and the form factor ${\cal G}(w)$
is generally expressed as the product of a normalization factor
${\cal G}(1)$ and a function, $g_D(w)$, constrained by dispersion
relations \cite{grinstein}.

The strategy to extract $\rm {\cal G}(1)|V_{cb}|$ is identical to
that used for the $B\dec D^{\star} \ell \nu$ decay. However, in
this case there is no suppression of $1/m_Q$ (i.e. no Luke
theorem) and corrections and QCD effects on ${\cal G}(1)$ are
calculated with less accuracy than ${\cal F}(1)$
\cite{NeubertPLB26491} \cite{LigeNirNewPRD4994}. Moreover,
${d\Gamma_D}/{dw}$ is more heavily suppressed near $w=1$ than
${d\Gamma_{D^*}}/{dw}$ due to the helicity mismatch between
initial and final states. This channel is much more challenging
also from the experimental point of view as it is hard to isolate
from the dominant background $ \rm B\dec D^\star \ell \nu$ as well
as from fake $\rm D$-$\ell$ combinations.  Thus, the extraction of
$\rm |V_{cb}|$ from this channel is less precise than the one from
the $\rm B\dec D^\star \ell \nu$ decay. Nevertheless,  the $B\dec
D \ell \nu$ channel provides a consistency check and allows a test
of heavy-quark symmetry \cite{LigeNirNewPRD4994} through the
measurement of the form factor ${\cal G}(w)$, as  HQET predicts
the ratio ${\cal G}(w)/{\cal F}(w)$ to be very close to one.

Belle \cite{belle-dplnu} and ALEPH \cite{ALEPH_vcb} studied the
$\rm \bar{B}^0\dec D^+ \ell^- \bar{\nu}$ channel, while CLEO
\cite{cleo-dplnu} studied both $\rm B^+\dec D^0 \ell^+ \bar{\nu}$
and $\rm \bar{B^0}\dec D^+ \ell^- \bar{\nu}$ decays. The published
results are shown in Table~\ref{t:D}, as well as the results
scaled to common inputs. Averaging the latter data, using the
procedure of \cite{lepvcb}, Averaging \cite{lepvcb} the data in
Table~\ref{t:D}, using the procedure  of \cite{lepvcb}, we get
${\cal G}(1) |V_{cb}| = (41.3  \pm  4.0)\times 10^{-3}$ and
$\rho_D^2 = 1.19 \pm 0.19,$ where $\rho_D^2$ is the slope of the
form-factor given in \cite{CLN} at zero recoil.

\begin{table}
\begin{center}
\begin{tabular}{|l|cc|} \hline
experiment   &  ${\cal G}(1)|V_{cb}|(\times 10^{3})$ & $\rho^2_D$ \\
\hline
Published values &  & \\
 ALEPH       &  31.1$\pm$  9.9$\pm$ 8.6 & 0.20$\pm$ 0.98$\pm$ 0.50 \\
 Belle       &  41.1$\pm$  4.4$\pm$ 5.2 & 1.12$\pm$ 0.22$\pm$ 0.14\\
 CLEO        &  44.4$\pm$  5.8$\pm$ 3.7 & 1.27$\pm$ 1.25$\pm$ 0.14 \\
\hline \hline
Scaled values & & \\
 ALEPH       &  37.7$\pm$  9.9$\pm$ 6.5 & 0.90 $\pm$ 0.98$\pm$ 0.38  \\
 Belle       &  41.2$\pm$  4.4$\pm$ 5.1 & 1.12$\pm$ 0.22$\pm$ 0.14 \\
 CLEO        &  44.6$\pm$  5.8$\pm$ 3.5 & 1.27$\pm$ 0.25$\pm$ 0.14 \\
\hline
World average &  41.3 $\pm$ 2.9$\pm$ 2.7 & 1.19$\pm$ 0.15$\pm$ 0.12 \\
\hline
\end{tabular}
\end{center}
\caption{Experimental results before and after the correction to
common inputs and world average. $\rho_D^2$ is the slope of the
form-factor given in \cite{CLN} at zero recoil. \label{t:D}}
\end{table}


Theoretical predictions for ${\cal G}(1)$ are consistent:
$1.03 \pm 0.07 $ \cite{ScoraIsgurPRD5295}, and $0.98 \pm 0.07 $
\cite{LigeNirNewPRD4994}.
A 
quenched lattice calculation gives ${\cal G}(1)=1.058
^{+0.020}_{-0.017}$ \cite{JLQCDPRL8299}, where the errors does not
include the uncertainties induced by the quenching approximation
and lattice spacing. Using ${\cal G}(1)=1.0\pm\ 0.07$, we get
$|V_{cb}| = (41.3 \pm 4.0_{exp} \pm 2.9_{theo}) \times 10^{-3},$
consistent with the value extracted from $\rm B\dec D^{\star} \ell
\nu$ decay, but with a larger uncertainty.

The experiments have also measured the differential decay rate
distribution to extract the ratio ${\cal G}(w)/ {\cal F}(w)$. The
data are compatible with a universal from factor as predicted by
HQET. From the measured values of ${\cal G}(1)|V_{cb}|$ and
${\cal F}(1)|V_{cb}|$, we get ${\cal G}(1)/{\cal F}(1) = 1.08 \pm 0.09,$
consistent with the form-factor values we used.

\section{$\vcb$ determination from inclusive B semileptonic decays}

Alternatively, \vcb\ can be extracted from the inclusive branching
fraction for semileptonic $b$ hadron decays ${\cal B} ( B \to X_c
\ell \nu )$ \cite{142}, \cite{143}. Several studies have shown
that the spectator model decay rate is the leading term in a well
defined expansion controlled by the parameter $\Lambda
_{QCD}/m_b$. Non-perturbative corrections to this leading
approximation arise only to order $1/m_b^2$. The key issue in this
approach is the ability to separate non-perturbative corrections,
that can be expressed as a series in powers of $1/m_b$, and
perturbative corrections, expressed in powers of $\alpha _S$.
Quark-hadron duality is an important {\it ab initio} assumption in
these calculations. While several authors \cite{bigiduality} argue
that this ansatz does not introduce appreciable errors as they
expect that duality violations affect the semileptonic width only
in high powers of the non-perturbative expansion, other authors
recognize that an unknown correction may be associated with this
assumption \cite{buchalla}. Arguments supporting a possible
sizeable source of errors related to the assumption of
quark-hadron duality have been proposed \cite{nathan}. This issue
needs to be resolved with further measurements. At present, no
explicit additional error has been added to account for a possible
quark-hadron duality violations.

The coefficients of the $1/m_b$ power terms that are valid through
order $1/m_b^2$ include four parameters: the expectation value of
the kinetic operator, corresponding to the average of the square
of the heavy quark momentum inside the hadron, the expectation
value of the chromomagnetic operator and the heavy quark masses
($m_b$ and $m_c$). The expectation value of the kinetic operator
is introduced in the literature as $-\lambda _1$ \cite{gremm-kap},
\cite{falk} or $\mu_{\pi}^2$ \cite{142} \cite{143}, whereas the
expectation value of the chromomagnetic operator is defined as
$\lambda _2$ \cite{gremm-kap}, \cite{falk} or $\mu_G^2$ \cite{142}
\cite{143}. The two notations reflect a difference in the approach
used to handle the energy scale $\mu$ used to separate the long
distance from the short distance physics. HQET is most commonly
renormalized in a mass-independent scheme, thus making the quark
masses 
the pole masses of the underlying theory (QCD).
The second group of authors prefer the definition of
the non-perturbative operators using a mass scale $\mu \approx 1$
GeV.

The equation for the semileptonic width according to the first set
of authors can be found in Ref. \cite{adam1}, that has been used
to extract $\rm |V_{cb}|$ from the semileptonic branching
fraction:
\begin{eqnarray} \Gamma_{sl} = & \frac{G^2_F \vert V_{cb} \vert^2
M_B^5}{192 \pi^3}0.3689 [1 - 1.54 \frac{\alpha_s}{\pi} - 1.43
\beta_0 \frac{\alpha_s^2}{\pi^2}
 - 1.648 \frac{\bar \Lambda}{M_B}(1 -
0.87 \frac{\alpha_s}{\pi} ) - 0.946\frac{\bar \Lambda^2}{M_B^2}
 -3.185 \frac{\lambda_1}{M_B^2}  \nonumber \\
&+ 0.02 \frac{\lambda_2}{M_B^2} -0.298 \frac{\bar
\Lambda^3}{M_B^3} - 3.28 \frac{\bar \Lambda \lambda_1}{M_B^3}
+10.47 \frac{\bar \Lambda \lambda_2}{M_B^3}
 -6.153 \frac{\rho_1}{M_B^3}  +7.482 \frac{\rho_2}{M_B^3} \nonumber \\
&- 7.4 \frac {{\cal T}_1}{M_B^3} + 1.491 \frac {{\cal T}_2}{M_B^3}
-10.41 \frac {{\cal T}_3}{M_B^3} -7.482 \frac {{\cal T}_4}{M_B^3}
+ {\cal O}(1/M^4_B)]\ . \label{eq:vcb}
\end{eqnarray}
 Eq.~\ref{eq:vcb}, in the $\overline
{MS}$ scheme, is calculated to order $1/M_B^3$ and $\beta_0
\alpha_s^2$. In order to derive this equation, the quark masses
are related to the corresponding meson masses through
\cite{neubert-vcb}:
\begin{equation}
m_Q = {\overline{M_M}}-{\overline{\Lambda}}+\frac{\lambda_{1}}{2
m_Q},
\end{equation}
where $m_Q$ is the heavy quark mass,
${\overline{M_M}}$ is the spin averaged heavy meson mass,
(${\overline{M_B}} = 5.313 GeV/c^2$ and ${\overline{M_D}} = 1.975
GeV/c^2$). The constant coefficients $\rho_1$, $\rho_2$, $\tau_1$,
$\tau_2$, $\tau_3$ and $\tau_4$ are form factors of different
contributions that arise to order $1/m_b^3$ and are described in
more detail in Ref. \cite{gremm-kap}.

The corresponding equation for the semileptonic width  in the
second approach is \cite{uraltsev}:
\begin{eqnarray}
\Gamma_{sl} & = &\frac{G_F^2 m_b^5}{192 \pi^3}\cdot [
z\left(\frac{m_c^2}{m_b^2}\right)
\left[1-a_1\frac{\alpha_s}{\pi}-a_2(\frac{\alpha_s}{\pi})^2
+...\right]\cdot
\left(1-\frac{\mu_{\pi}^2-\mu_G^2}{2m_b^2}\right)\nonumber \\
 & &  -( 1+b_1\frac{\alpha_s}{\pi}+...)\cdot
(1-\frac{m_c^2}{m_b^2})^4
\frac{2\mu_G^2}{m_b^2}-d\frac{\rho_D^3}{m_b^3}],
\label{eq:vcb-kolya}
\end{eqnarray}
where $m_b(\mu)$ and $m_c(\mu)$ are short scale quark masses, $z$
is a known parton phase space factor dependent upon $m_c^2/m_b^2$,
$a_i$ and $b_i$ parameters are coefficient of the perturbative
expansion \cite{uraltsev}: they, as well as the parameter $d$, are
function of $(m_c^2/m_b^2)$. The parameter $\rho _D$ gives the
strength of the so called `Darwin term', the dominant $1/m_b^3$
correction in this approach. The short distance mass $m_b(\mu \sim
1 GeV)$ has been evaluated with different techniques. The central
value is rather consistent and the errors vary from 30 to 110 MeV
\cite{hoang-ckmw}.  Thus we consider $\rm m_b(\mu\sim 1 \rm GeV) =
4.58 \pm 0.09\ GeV$ the best representation of our present
knowledge.

\subsection{HQE and moments in semileptonic decays and $b\rightarrow s\gamma$}

Experimental determinations of the HQE parameters are important in
several respects. In particular, redundant determinations of these
parameters may uncover inconsistencies, or point to violation of
some important assumptions inherent in these calculations. The
parameter $\lambda _2$ can be extracted from the $\rm B^{\star}-B$
mass splitting and found to be $\lambda _2= 0.128\pm 0.010$
GeV$^2$ \cite{gremm-kap},whereas the other parameters need more
elaborate measurements.

The first stage of this experimental program has been completed
recently. The CLEO collaboration has measured the shape of the
photon spectrum in $\rm b\rightarrow s \gamma$ inclusive decays.
Its first moment, giving the average energy of the $\gamma$
emitted in this transition, is related to the $b$ quark mass. In
the formalism of Ref. \cite{adam1} this corresponds to the
measurement of the parameter $\overline{\Lambda}=0.35\pm 0.07\pm
0.10$ GeV.

The parameter $\lambda _1$ is determined experimentally through a
measurement of the first moment of the mass $M_X$ of the hadronic
system recoiling against the $\ell$-$\bar{\nu}$ pair. This CLEO
measurement takes advantage of the ability of reconstructing the
$\nu$ 4-momentum with high efficiency and resolution, by virtue of
the hermeticity of the detector and the simplicity of the initial
state. The relationship between the first moment of $M_X$, defined
as ${\cal M}_1\equiv<M_X^2-\bar{M}_D^2>/M_B^2$, and the parameters
$\overline{\Lambda}$ and $\lambda _1$ is given in \cite{aref}.
\begin{eqnarray}
    {\cal M}_1\equiv\frac{\langle M_X^2 - \bar M_D^2 \rangle}{M_B^2} = &
[0.0272 \frac{\alpha_s}{\pi} + 0.058 \beta_0
\frac{\alpha_s^2}{\pi^2} + 0.207 \frac{\bar \Lambda}{\bar
M_B}(1+0.43 \frac{\alpha_s}{\pi}) + 0.193 \frac{\bar
\Lambda^2}{\bar M_B^2} + 1.38 \frac{\lambda_1}{\bar M_B^2} \nonumber \\
& + 0.203 \frac{\lambda_2}{\bar M_B^2} +0.19 \frac{\bar
\Lambda^3}{\bar M_B^3} + 3.2 \frac{\bar \Lambda \lambda_1}{\bar
M_B^3}
+1.4 \frac{\bar \Lambda \lambda_2}{\bar M_B^3} +4.3 \frac{\rho_1}{\bar M_B^3}\nonumber \\
&   -0.56 \frac{\rho_2}{\bar M_B^3} + 2.0 \frac {{\cal T}_1}{\bar
M_B^3} + 1.8 \frac {{\cal T}_2}{\bar M_B^3} + 1.7 \frac {{\cal
T}_3}{\bar M_B^3} + 0.91 \frac {{\cal T}_4}{\bar M_B^3} + {\cal
O}(1/\bar M^4_B)]. \label{eq:hadmom1}
\end{eqnarray}

The measured value for $<M_X^2-M_D^2>$ is $0.251 \pm 0.066\ {\rm
GeV}^2$. This constraint, combined with the measurement of the
mean photon energy in $b\rightarrow s\gamma$, implies a value of
$\lambda_1 = -0.24 \pm 0.11\ {\rm GeV}^2$, to order $1/M_B^3$ and
$\beta_0 \alpha_s^2$ in $(\overline{MS})$. The quoted theoretical
uncertainty of 2\% accounts for the $1/M_B^3$ and $\alpha_s$
uncertainties, but not for possible violations of quark-hadron
duality. Thus more conservative uncertainties have been used in
the literature for our present knowledge of this parameter
\cite{adam-review}. The next step involves a determination of the
HQE parameters with independent measurements, for example, the
moments of the lepton energy spectrum \cite{voloshin},
\cite{zoltan}. Preliminary data show contradictory implications
\cite{zoltan}, \cite{ron-vancouver}; this issue should be settled
soon with definitive results from several different experiments.

\subsection{Experimental determination of the semileptonic branching fraction}

The value of ${\cal B}( \rm B \to X_c \ell \nu )$ has been measured
both at the $\Upsilon(4S)$ and LEP.

The most recent CLEO data, published in 1996 and based on a subset
of the data sample accumulated now, obtains this branching
fraction using a lepton tagged sample \cite{CLEOl}. In this
approach, a di-lepton sample is studied, and the charge
correlation between the two leptons allows to disentangle leptons
coming from the direct decay $ B\rightarrow X_c \ell \nu$ and the
dominant background at low lepton momenta, the cascade decay $\rm
B\rightarrow X_c\rightarrow X_s\ell \nu$. This method, pioneered
by the ARGUS collaboration \cite{argus-dilept} allows the electron
spectrum from $\rm B\rightarrow X_c \ell \nu$ to be measured down
to 0.6 GeV/c. Thus, it allows to reduce the model dependence of
the extracted semileptonic branching fraction very substantially.
They obtain ${\cal B}(\rm B\rightarrow X_c e\nu) = (10.49 \pm
0.17\pm 0.43) \%$. The systematic error (4\%) is dominated by
experimental uncertainties. Lepton identification efficiency, fake
rate determination and tracking efficiencies contribute to 3\% of
this overall error. The remaining error is a sum of several small
corrections associated to the uncertainty in the mixing parameter,
and additional background estimates \cite{CLEOl}.

Combining $\rm \Upsilon(4S)$ results \cite{PDG2002}, we obtain: $
{\cal B} (\rm b \to X \ell \nu) = (10.38 \pm 0.32) \% .$ Using
$\tau_{B^+}$, $\tau_{B^0}$ \cite{PDG2002},
$f_{+-}/f_{00} = 1.04 \pm 0.08 $ \cite{CLEO-ch-neu} and
subtracting ${\cal B}\rm (b \to u \ell \nu) =  (0.17 \pm 0.05)\%$,
we get:
$  {\cal B} (b \to X_c \ell \nu) = (10.21 \pm 0.32) \%$ and
$\rm \Gamma (b \to X_c \ell \nu) = (0.419 \pm 0.013 \pm 0.003)
\times 10^{-10}\ MeV,$
where $0.003 \times 10^{-10}\ {\rm MeV}$ includes the
uncertainties from ${\cal B}(\rm b \to u \ell \nu)$ and the model
dependence, correlated with LEP.

At LEP, $\rm B^0_d$, $\rm B^-$, $\rm B_s$ and b-baryon are produced,
so the measured inclusive semileptonic branching ratio is an average
over the different hadron species.  Assuming that the semileptonic
widths of all b-hadrons are equal, the following relation holds:
\begin{eqnarray}
{\cal B}( b \to X_c l \nu )_{LEP}   =
       & f_{B^0} \frac{\rm \Gamma( B^0 \to X_c l\nu )}{\Gamma(B^0)} +
      f_{B^-} \frac{\rm \Gamma( B^- \to X_c l\nu )}{\Gamma(B^-)} +
      f_{B_s} \frac{\rm \Gamma( B_s \to X_c l\nu )}{\Gamma(B_s)} +
      f_{\Lambda_b} \frac{\rm \Gamma( \Lambda_b \to X_c l\nu )}
                               {\Gamma(\Lambda_b)} = \nonumber \\
     & \Gamma({\rm B \to X_c \ell \nu}) \tau_{\rm b}  \label{eq:bri}
\end{eqnarray}
where $\tau_{\rm b}$
is the average b-hadron lifetime.
Taking into account the present precision of LEP measurements of b-baryon
semileptonic branching ratios and lifetimes, the estimate uncertainty
for a possible difference for the width of b-baryons is $0.13\%$.

At LEP, ${\cal B}(\rm b \to X \ell \nu)$ is measured with
dedicated analyses \cite{aleph}, \cite{D}, \cite{L}, \cite{O},
summarized in (Table~\ref{t:inc}). The average LEP value for
${\cal B} (\rm b \to X \ell \nu) = (10.59 \pm 0.09 \pm 0.30)\% $
is taken from a fit \cite{EWWG}, which combines the semileptonic
branching ratios, the $\rm B^0-\bar{B^0}$ mixing parameter
$\bar{\chi_b}$, and $\rm R_b =\Gamma(Z\to b\bar{b})/\Gamma(Z\to
had)$.

Ref. \cite{aleph} shows that the main contribution to the
modelling error is the uncertainty in the composition of the
semileptonic width including the narrow, wide and non-resonant
$\rm D^{\star\star}$ states. $\rm B_s$ and b-baryons are about
20\% of the total signal and their contribution to the uncertainty
of the spectrum is small.
In this average, we use the modelling error quoted by \cite{aleph}
rather than the error from the combined fit, as the ALEPH
procedure is based on more recent information. The dominant errors
in the combined branching fraction are the modelling of
semileptonic decays (2.6\%) and the detector related items
(1.3\%).

\begin{table}
\begin{center}
\begin{tabular}{|l|c|} \hline
 Experiment &  ${\cal B}(b\dec\ell\nu) \%$  \\
 \hline
 ALEPH      &  10.70 $\pm$ 0.10 $\pm$ 0.23 $\pm$ 0.26 \\
 DELPHI     &  10.70 $\pm$ 0.08 $\pm$ 0.21 $\pm^{+0.44}_{-0.30}$ \\
 L3         &  10.85 $\pm$ 0.12 $\pm$ 0.38 $\pm$ 0.26 \\
 L3 (double-tag)         &  10.16 $\pm$ 0.13 $\pm$ 0.20 $\pm$ 0.22 \\
 OPAL       &  10.83 $\pm$ 0.10 $\pm$ 0.20 ${\pm}^{+0.20}_{-0.13}$ \\ \hline
 LEP Average & 10.59 $\pm$ 0.09 $\pm$ 0.15 $\pm$ 0.26 \\ \hline
\end{tabular}
\end{center}
\caption{$ {\cal B} (\rm b \to \ell )$ measurement from LEP and their average.
The errors quoted reflect statistical, systematic and modelling
uncertainties respectively. \label{t:inc}}
\end{table}

Subtracting ${\cal B}\rm (b \to u \ell \nu)$ from the LEP semileptonic
branching fraction, we get:
$ {\cal B} (b \to X_c \ell \nu) = (10.42 \pm 0.34) \% ,$
and using $\tau_b$ \cite{PDG2002}:
$\rm \Gamma (b \to X_c \ell \nu) = (0.439 \pm 0.010 \pm 0.011)\times 10^{-10}\ {\rm MeV},$
where the systematic error $0.011\times\ 10^{-10} {\rm MeV}$
reflects the ${\cal B}\rm  (b \to u \ell \nu)$ uncertainty and the
model dependence, correlated with the $\rm \Upsilon(4S)$ result.

Combining the LEP and the $\rm \Upsilon(4S)$ semileptonic widths,
we get: $\rm \Gamma (b \to X_c \ell \nu) = (0.43 \pm 0.01)\times
10^{-10}\ {\rm  MeV},$ which is used in the formula of Ref.
\cite{gremm-kap}
to get:
$${\rm |V_{cb}|_{incl}} = (40.4 \pm 0.5_{exp} \pm
0.5_{\lambda_1,\overline{\Lambda}}\pm 0.8_{theo}) \times
10^{-3}.$$ where the first error is experimental, the second from
the measured value of $\lambda_1$ and $\overline{\Lambda}$,
assumed to be universal up to higher orders. The third error is
from $1/m_b^3$ corrections and from the ambiguity in the
$\alpha_s$ scale definition. The error on the average b-hadron
lifetime is assumed to be uncorrelated with the error on the
semileptonic branching ratio.

\section{Conclusions}

The values of $\vcb$ obtained both from the inclusive and exclusive method
agree within errors.
The value of \vcb\ obtained from the analysis of the $\rm B\rightarrow
D^{\star}\ell \nu$ decay is:
\begin{equation}
\rm |V_{cb}|_{exclusive} = (42.1 \pm 1.1_{exp} \pm 1.9_{theo})
\times 10^{-3}
\end{equation}
where the first error is experimental and the second error is from
the $1/m_b^2$ corrections to ${\cal F}(1)$. The value of $\vcb$,
obtained from inclusive semileptonic branching fractions is:
\begin{equation}
\rm |V_{cb}|_{incl} = (40.4 \pm 0.5_{exp} \pm
0.5_{\lambda_1,\overline{\Lambda}}\pm 0.8_{theo}) \times 10^{-3},
\end{equation}
where the first error is experimental, the second error is from
the measured values of $\lambda_1$, and $\overline{\Lambda}$,
assumed to be universal up to higher orders, and the last from
$1/m_b^3$ corrections and $\alpha_s$. Non-quantified uncertainties
are associated with a possible quark-hadron duality violation. For
this reason, we chose not to average the two numbers.

While experimental errors have reached $2.7\%$ and $1.2\%$ levels
respectively, the dominant uncertainties remain of theoretical
origin. Thus an unambiguous statistical treatment of these
uncertainties is very difficult. High precision tests of HQET,
checks on possible violations of quark-hadron duality in
semileptonic decays, experimental determination of $m_b$,
$m_b-m_c$ and $\mu_{\pi}^2$ are needed to complete this
challenging experimental program.

\section{Acknowledgements}
We would like to acknowledge useful contributions from K. Ecklund,
J.H. Ki, K. Moenig and P. Roudeau. Moreover we would like to thank
C. Bauer, I.I. Bigi, A. Falk, Z. Ligeti, A. Kronfeld, T. Mannel,
S. Stone and N.G. Uraltsev for interesting discussions. This work
was supported in part by National Science Foundation.

\newpage

\end{document}